\documentstyle[bezier,psfig,12pt]{article} 
       
     \makeatletter 

 \newcommand{\integer}{{ 
  \setbox0\hbox{\m@th$\fam\sffam Z$} 
  \setbox1\hbox{\rm\kern.05\wd0 
                \rlap{\vrule height.93\ht0 depth-.75\ht0 width.056\wd0 }%
                \kern-.13\wd0 \copy0 \kern-.6\wd0 \copy0 \kern-.1\wd0 
                \llap{\vrule height.25\ht0 depth\z@ width.056\wd0}%
                \kern.05\wd0} 
  \mathchoice{\copy1}{\copy1}{\mit Z\mkern-8mu Z}{\mit Z\mkern-7.5mu Z} }} 
\def\@@insvline#1#2{{\setbox0\hbox{\m@th$#1\rm I$} 
  \rlap{\m@th$#1 \mkern 5mu 
  \vrule height.92\ht0 depth-.05\ht0 width.09\ht0 $} 
  {\rm #2} }} \newcommand{\rat}{\mathpalette\@@insvline{Q}} 
 
\newcommand{\complex}{\mathpalette\@@insvline{C}}

\makeatother

    \def\be{\begin{equation}}
    \def\ee{\end{equation}}
    \def\bea{\begin{eqnarray}}
    \def\eea{\end{eqnarray}}
    \def\f{\frac}

       \def\bc{\begin{center}}
     \def\ec{\end{center}} 
      \def\p{\partial} 
        
   \def\ignore#1{\par}

\begin{document} 

\title{The Dirac-Maxwell Equations with Cylindrical Symmetry}

\author{H.S. Booth and C.J. Radford}
\date{July 29, 1996}
\maketitle

\vspace{5cm}

\begin{abstract}
A reduction of the Dirac-Maxwell equations in the case of
static cylindrical symmetry is performed.
The behaviour of the resulting system of o.d.e.s 
is examined analytically and numerical solutions presented.
There are two classes of solutions.
\begin{itemize}
   \item The first type of solution is a Dirac field surrounding a
 charged ``wire.'' The Dirac field is highly 
localised, being concentrated in cylindrical shells about the wire.
A comparison with the usual linearized theory demonstrates that this
localisation is entirely due to the non-linearities in the
equations which result from the inclusion of the ``self-field''.
   \item  The second class of solutions have the electrostatic
potential finite along the axis of symmetry but unbounded at large
distances from the axis.
 \end{itemize} 
\end{abstract}
\pagebreak

\section{Introduction}

This paper is concerned with the system of partial differential equations known 
as the Dirac-Maxwell equations; the Dirac field describing spin-half matter 
and the Maxwell field mediating the electromagnetic interactions of the Dirac 
field. In an earlier paper, \cite{rad}, one of us (C.J.R.) used the two-spinor 
formalism to rewrite the equations in a novel way.
In fact the electromagnetic potential can be eliminated from the equations
altogether. 
(For further background into the two-spinor formulation of the
Dirac equation, see \cite{pen}).

In \cite{rad} a reduction 
of the equations was performed in the static case (see section 2, below) and a
 signifigantly simplified set of (nonlinear) p.d.e.s presented. Finally, after 
imposing spherical symmetry, the equations were solved (numerically, at least).
 Unusual features to emerge from that investigation were the existence of highly
 compact objects and a magnetic monopole. In the present paper we will solve 
(numerically) the cylindrically symmetric, static Dirac-Maxwell equations. In 
this case we find no monopole-like solutions. Perhaps this is not unexpected, 
since there is slightly 
more regularity in a cylindrical system as opposed to a spherical system --
 $\ln(r)$ behaviour as distinct from $1/r$ behaviour. However, we do find highly
 localised solutions -- cylindrical shells about a central charged ``wire''. 
When we compare this to the Dirac field solved in an external
potential of the same type, no such localisation is apparent.

The cylindrical reduction is also distinguished from the spherical case by the 
existence of unbounded solutions -- solutions in which  
the Maxwell potential is unbounded at large 
distances from the symmetry axis. These solutions are also unbounded
in the total charge (per unit length of ``wire'').

Very few explicit solutions to the full Dirac-Maxwell equations are known and 
all of these are numerical solutions (or partly so) -- see \cite{rad}, 
\cite{wak} and \cite{lis}.
Work on the existence theory has progressed steadily, with some important recent
 work (see \cite{est1} and \cite{est2}) providing local existence for soliton 
like solutions.

 The solutions in \cite{rad} and those presented here show that these equations are 
capable of representing highly localised structured, entities. Such behaviour 
is not even hinted at when one examines the usual linearised theory (see
section{\bf 5.2}), in which 
one ignores the Dirac current as a source for the Maxwell field.

\section{The Static Dirac-Maxwell Equations}

The Dirac-Maxwell equations in standard notation are,
\be
\gamma^{\alpha}(\p_{\alpha}-ieA_{\alpha})\psi + im\psi  = 0,\;\;\;\;\mbox{where}\;\;\;\;\alpha=0,\ldots,3   \label{eq:dirac4}
\ee
\be
\mbox{and}\;\;\;\;\p^{\alpha}F_{\alpha\beta} =  -4\pi ej_{\beta},\label{eq:max}
\ee
\be
\mbox{where}\;\;\;\;F_{\alpha\beta}=A_{\beta,\alpha}-A_{\alpha , \beta} 
 \nonumber \label{eq:fab}
\ee
with the current density given by
\be
j_{\alpha}=\overline{\psi}\gamma_{\alpha}\psi.     \nonumber  \label{eq:current}
\ee
Following [cjr] we use the $\gamma _{5}$ diagonal or van der Waerden description
 (2-spinor notation, see~\cite{pen}). The Dirac 4-spinor (or bispinor) is 
\[\psi = \left( \begin{array}{c} u_{A} \\ 
                              \bar{v}^{\dot{B}} \end{array} \right)\;\;\;\;\;
\mbox{and} \;\;\;\;\;\overline{\psi} = \left( v^{B}, \bar{u}_{\dot{A}}\right).\]
The Dirac equations are
\bea
(\p^{A\dot{A}}-ieA^{A\dot{A}})u_{A}+\f{im}{\sqrt{2}}\bar{v}^{\dot{A}} & = & 0  \nonumber \\
(\p^{A\dot{A}}+ieA^{A\dot{A}})v_{A}+\f{im}{\sqrt{2}}\bar{u}^{\dot{A}} & = & 0.   \label{eq:dirac2} 
\eea
In [cjr] it was shown that these equations can be solved for the
 electromagnetic potential, provided the 2-spinors meet a non-degeneracy 
condition, $u^{A}v_{A}\neq 0$. The result is,
\be
A^{A\dot{A}}=\f{i}{e(u^{C}v_{C})}\left(v^{A}\p^{B\dot{A}}u_{B}+u^{A}\p^{B\dot{A}}v_{B}+\f{im}{\sqrt{2}}(u^{A}\bar{u}^{\dot{A}}+v^{A}\bar{v}^{\dot{A}})\right). \label{eq:AAA}
\ee
We are interested in the classical Dirac-Maxwell equations, so we will also 
require that $A^{A\dot{A}}$ is a real vector field (see [cjr]). This leads to 
the following first order differential equations for $u_A$ and $v_A$ (``reality
 conditions"):
\bea
\p^{A\dot{A}}(u_{A}\bar{u}_{\dot{A}})  & = & - \f{im}{\sqrt{2}}(u^{C}v_{C}-\bar{u}^{\dot{C}}\bar{v}_{\dot{C}})   \nonumber  \\
\p^{A\dot{A}}(v_{A}\bar{v}_{\dot{A}})  & = &  \f{im}{\sqrt{2}}(u^{C}v_{C}-\bar{u}^{
\dot{C}}\bar{v}_{\dot{C}})   \nonumber  \\
u_{A}\p^{A\dot{A}}\bar{v}_{\dot{A}}-\bar{v}_{\dot{A}}\p^{A\dot{A}}u_{A} & = & 0. \label{eq:reals}
\eea

The expression for the potential (6), the reality conditions (7) and the Maxwell
 equations (2), now constitute the full set of nonlinear partial differential 
equations for the Dirac-Maxwell system. We still have, of course, the $U(1)$ 
gauge freedom,
\[\begin{array}{l}
u_A\rightarrow e^{i\zeta}u_{A}\\
v^A\rightarrow e^{-i\zeta}v^A\\
A_{\alpha}\rightarrow A_{\alpha}+\f{1}{e}\p_{\alpha}\zeta,\;\;\;\;\,
\mbox{a direct consequence of (6).}
\end{array}\]

Following [cjr] we now impose the {\em static condition}: there exists a Lorentz
 frame in which there is no current ``flow", $j^{\alpha}=\delta^{\alpha}_{0}j^0$
. Under this condition we have
\be
v^{A}=e^{i\chi}\sqrt{2}\sigma^{0A\dot{A}}\bar{u}_{\dot{A}} \ ,
\ee
where $\chi$ is an arbitrary real function
and $\sigma^{\alpha A\dot{A}}$ are the three Pauli matrices with the $2\times2$
identity matrix as the zeroth $\sigma$ matrix (van der Waerden symbols).

The gauge can be fixed by the choice
\bea
u^{0} & = & X e^{\f{i}{2}(\chi+\eta)}  \nonumber \\
u^{1} & = & Y e^{\f{i}{2}(\chi-\eta)} \ ,  \label{eq:stspins}
\eea
where $\eta$, $X$ and $Y$ are real functions.

The {\em static} Dirac-Maxwell equations can now be written down, we follow 
[cjr] and write them in 3-vector notation. This is done by first introducing 
vectors ${\mbox {\boldmath $V$}}$ and ${\mbox {\boldmath $A$}}$,
\[\begin{array}{l}
             {\mbox {\boldmath $V$}}= (2XY\cos \eta,\,2XY\sin \eta,\,X^2-Y^2)\\
{\mbox {\boldmath $A$}}= (A^1,A^2,A^3),\,\mbox{where}\, A^j=\sigma^j_{B\dot{B}}
                                                             A^{B\dot{B}} \ . 
  \end{array}\]
Our Maxwell-Dirac equations are now given as follows. \\       
The electromagnetic potential is
\bea
A^0 & = & \f{m}{e}\cos \chi + \f{(X^2-Y^2)}{2e(X^2+Y^2)}\f{\p\eta}{\p t}
         +\f{({\mbox {\boldmath $\nabla$}}\chi ){\mbox {\boldmath $.$}}
{\mbox {\boldmath $V$}}}{2e(X^2+Y^2)} \nonumber \\
{\mbox {\boldmath $A$}} & = & \f{1}{2e(X^2+Y^2)}\left[\f{\p\chi}{\p t}
{\mbox {\boldmath $V$}}+(X^2-Y^2)
{\mbox {\boldmath $\nabla$}}\eta -{\mbox {\boldmath $\nabla$}}\times
{\mbox {\boldmath $V$}}\right].      \label{eq:pot}
\eea
The reality conditions are
 \bea
\f{\p}{\p t}(X^2+Y^2) & = & 0   \\
   {\mbox {\boldmath $\nabla$}}{\mbox {\boldmath $.$}}{\mbox {\boldmath $V$}}
   & = & -2m(X^2+Y^2)\sin \chi\\
\f{\p {\mbox {\boldmath $V$}}}{\p t}+\left( {\mbox {\boldmath $\nabla$}}\chi
\right) \times
{\mbox {\boldmath $V$}} & = &  {\boldmath 0}.
\eea
Together with the Maxwell equations for {\mbox {\boldmath $A$}} with current
 vector $j$, as above.

\section{Cylindrical Symmetry}

The Dirac field given by the 2-spinors $u_A$ and $v^A$ determines two gauge 
invariant null vectors 
$l^{\alpha}=\sigma^{\alpha A\dot{A}}u_A\bar{u}_{\dot{A}}$  and 
$n^{\alpha}=\sigma^{\alpha A\dot{A}}v_A\bar{v}_{\dot{A}}$. In fact,
\[(l^{\alpha})=\f{1}{\sqrt{2}}\left( X^2+Y^2,{\mbox {\boldmath $V$}}\right)
\,\;\;\mbox{and}\,\;\;(n^{\alpha})=\f{1}{\sqrt{2}}\left( X^2+Y^2,
-{\mbox {\boldmath $V$ }}\right),\]
assuming that the field is {\em static}.
Note that the current vector $j^{\alpha} = l^{\alpha} + n^{\alpha}$.
The first reality condition tells us that the zeroth component
of $l^{\alpha}$ and $n^{\alpha}$ are independent of time.
We also assume that $l^{\alpha}$ and $n^{\alpha}$ are independent
of $z$ in keeping with the assumption of cylindrical symmetry.  
Here $(\rho,\phi,z)$ are the cylindrical polar coordinates corresponding to
our original Cartesian coordinates. 

Applying these conditions one finds that $\eta =\phi$, $X=X(\rho)$ and 
$Y=Y(\rho)$. The second of the reality conditions then implies that 
$\chi =\chi (\rho)$. The third reality condition then gives 
\[(X^2-Y^2)\f{d\chi}{d\rho}=0.\]
If we take $\chi =constant$ then we find $A^0 =constant$, the Maxwell equation
then implies that $X=Y=0$. However, the non-degeneracy condition,
$u^{A}v_{A}\neq0$, excludes this possibility. So we assume $\chi '(\rho) \neq 0$
 from now on and consequently, $X=Y$.
 Our equation for the electromagnetic potential (\ref{eq:pot}), 
shows that the vector potential ${\mbox {\boldmath $A$}}$ vanishes. There is
 no magnetic monopole, as distinct from the spherical case.

The equations now simplify to the following ordinary differential equations:
\bea
A_{0}& = &\f{m}{e} \cos \chi + \f{1}{2e} \f{d \chi} {d \rho}\nonumber \\
\f{d}{d\rho} \left( \rho X^{2} \right) & = & -2m \rho X^{2} \sin \chi\nonumber\\
\f{d}{d\rho} \left( \rho \f{d A_{0}}{d\rho} \right) & = & 8\pi \sqrt{2} e X^{2}
 \rho.  
\eea

Introducing dimensionless variables
\bea
A  & = & \f{e}{m} A_{0}  \nonumber  \\
\hat{\rho}  & = & 2 m \rho  \nonumber  \\
Z  & =  & \f{8 \pi \sqrt{2} \rho e^{2}}{m^{2}} X^{2}.  \nonumber 
\eea
From here on we will refer to $\hat{\rho}$ as  $\rho$ for the sake of 
simplicity.
We also define a new dependent variable
\[ F = \rho \f{d A}{d \rho},  \]
so that we get a system of four first order equations
\bea
(a) \;\;\;\;\;\;\f{d \chi}{d \rho} & = & A - \cos \chi   \nonumber  \\
(b) \;\;\;\;\;\;\f{d F}{d \rho}   & = &  Z   \nonumber   \\
(c) \;\;\;\;\;\;\f{d A}{d \rho}   & = & \f{F}{\rho}    \nonumber  \\
(d) \;\;\;\;\;\;\f{d Z}{d \rho}   & = &  -Z \sin \chi.    \label{eq:4des} 
\eea

The equations can also be written as a fourth order equation in the dependent
\break
variable $ \chi $:

\be
\f{d^{2}}{d \rho^{2}} \left( \rho \left( \f{d^{2} \chi}{d \rho^{2}}-\sin \chi
\f{d \chi}{d \rho} \right) \right) +\f{d}{d \rho}\left( \rho \left(
\f{d^{2} \chi} {d \rho^{2}} -\sin \chi \f{d \chi}{d \rho} \right) \right)
\sin \chi = 0.                           \label{eq:1de}
\ee
Much of the qualitative behaviour of solutions to our system will be determined by the 
behaviour of the dependent variables in the vicinity of the two singular points
 $\rho =0$ and $\rho = \infty$.

\section{Behaviour Near $0$ And $\infty$}
The variable $Z$ is the charge density which we assume to be non-negative
(In fact, it is straightforward to show that $Z \ge 0 => Z>0$. See
proof of {\bf Lemma $2$}.)
$F$ is the integral of the charge $Z$ per unit
ring, that is, $F(\rho)-F(0)$ is the total charge within a radius $\rho$.
It is reasonable, therefore, to restrict our attention to solutions where
$F$ remains bounded as $\rho$ approaches $\infty$.
Much of the behaviour of the solution can be characterised
by the values of $F$ at either end of the domain.
This behaviour can best be summarized in the following two {\bf Lemmas}. 

\newtheorem{hilary1}{Lemma}
\begin{hilary1}
Suppose $(\chi,F,A,Z)$ is a solution to Equation(\ref{eq:4des}) on $I = 
(0,\rho_{1})$, for some $\rho_{1}, 0 < \rho_{1} < 1  $. Suppose also that 
 $Z \ge 0$ is continuous and bounded on $I$. Then,  
\begin{enumerate}
  \item[(i)] $F$ is $C^{1}$ on I and has a well defined, finite limit as        
 $\rho \rightarrow 0$. $Z$ has a well-defined limit as  $\rho \rightarrow 0$. 
  \item[(ii)] if $ F(0) \neq 0$ then $A$ is unbounded as $\rho \rightarrow 0$.
               In particular, $A=\Omega (\rho) \ln (\rho)$, where $\Omega$
               is $C^{2}$ and bounded on I,  $\Omega \rightarrow F(0)$ as
               $\rho \rightarrow 0$. Also, $\chi$ is bounded as $\rho 
\rightarrow 0$.
\end{enumerate}
\end{hilary1}
\newpage

{\bf Proof} 
\begin{enumerate}
   \item[(i)] Let $ 0 < \rho_{1} < \rho_{2} < 1 $.  \\
               From $(c)$ , $\f {d F}{d \rho} = Z$, so $F$ is $C^{1}$ and 
               $|F(\rho_{2}) - F(\rho_{1}) | \leq M(\rho_{2}-\rho_{1})$
               where $M = sup_{I} Z$. \\
            Letting $\rho_{1}, \rho_{2} \rightarrow 0$ shows that $F$
            has a well-defined limit. 
            Fixing $\rho_{2}$, and letting $\rho_{1} \rightarrow 0$ shows that
            this limit is finite. 
            A similar argument using (d);  
            $ \f {dZ} {d\rho} = - Z \sin {\chi} \leq M $, 
            shows that $Z$ has a well-defined limit.
    \item[(ii)]  Let $A = \Omega \ln \rho$. $A$ is $C^{2}$ on $I$ since,
             from (c), $\f {dA} {d \rho} = \f {F} {\rho} $;  
             therefore $\Omega$ is $C^{2}$. We also have 
\be
F = \rho \ln \rho \f {d \Omega} { d \rho} + \Omega   \label{eq:Omega1}
\ee
and
\be
\f {d} {d \rho} \left(  \rho (\ln \rho)^{2} \f {d \Omega} {d \rho} \right) = Z \ln \rho.
\label{eq:Omega2}
\ee
From (\ref{eq:Omega2}), (since $\ln \rho < 0$ on $I$),
\[ M \ln \rho \le \f{d}{d \rho} \left( \rho (\ln \rho)^{2} \f {d \Omega} 
{d \rho} \right) \le 0.  \]
Integrating, 
\[ M \left( \rho_{1}(\ln \rho_{1} -1)-\rho(\ln \rho -1) \right)
\le c_{1} - \rho (\ln \rho)^{2} \f {d \Omega} {d \rho}  \le 0, \]
where 
\[ c_{1} = \rho_{1} (\ln \rho_{1})^{2} \f{d \Omega}{d \rho}(\rho_{1}). \]
That is,
\[ \f { M \left( \rho_{1}(\ln \rho_{1} -1)-\rho(\ln \rho -1) \right)} {\ln \rho}\le \rho \ln \rho \f{d \Omega}{d \rho} \le \f {c_{1}}{\ln \rho}. \]
Letting $\rho \rightarrow 0$, we see that $\rho \ln \rho \f{d \Omega}{d \rho}
\rightarrow 0$. Hence, from (\ref{eq:Omega1}) we see that 
$\Omega \rightarrow F(0)$ as $\rho \rightarrow 0$. 

Now, since $\Omega$ is bounded on $I$ 
put $c_{2} \le \Omega \le c_{3}$, for some constants $c_{2}$ and $c_{3}$.
From (a),
\[ A -1 \le \f{d \chi}{d \rho} \le A + 1.\]
Thus,
\[ c_{2} \ln \rho -1 \le \f{d \chi}{d \rho} \le c_{3} \ln \rho +1.  \]
Integrating, we have
\bea
\chi(\rho)  & \le  & 
\chi(\rho_{1})  +  
c_{2}\left(\rho_{1}(\ln\rho_{1}-1)-\rho(\ln\rho-1)\right)-(\rho_{1}-\rho)
\nonumber  \\
\chi(\rho) & \ge & \chi(\rho_{1})+
c_{3}\left(\rho_{1}(\ln\rho_{1}-1)-\rho(\ln\rho-1)\right)+(\rho_{1} -\rho), 
\nonumber  
\eea
which bounds $\chi(\rho)$ as $\rho \rightarrow 0$.
\begin{flushright}  $\Box$  \end{flushright}

\end{enumerate}
\newtheorem{hilary2}[hilary1]{Lemma}
\begin{hilary2}
Suppose $(\chi,F,A,Z)$ is a solution to Equation(\ref{eq:4des}) on $\rho \in
(0,\infty)$. Suppose also that $Z \geq 0 $ with $F$ continous and bounded on the
interval. 
Then
\begin{enumerate}
  \item[(i)]If $F(\rho_{1}) \ge 0$ for some $\rho_{1}\in \left[0,\infty\right)$,
 then $\chi \rightarrow \infty$, $A \rightarrow \infty$, and $F \rightarrow
\infty$ as $\rho \rightarrow \infty$.
    \item[(ii)]  If $F < 0 $ on $(0,\infty)$ then 
$F \rightarrow 0$ as $\rho \rightarrow \infty$.
In addition, if $A$ and $Z$ have well-defined limits as $\rho \rightarrow
 \infty$ then $Z \rightarrow 0$ and $A \rightarrow A_{\infty}$
as $\rho \rightarrow \infty$, with $-1 \le A_{\infty} \le 1$.
\end{enumerate}
\end{hilary2}
{\bf Proof}
\begin{enumerate}
 \item[(i)]
We firstly show that, given the nondegeneracy condition, $u^{A}v_{A}\neq0$ a.e.,
that $Z \ge 0 => Z>0$
for all $\rho \in I$.
Suppose $Z(\rho_{1})=0$ for some $\rho_{1}\in I$.
Now $\f{d}{d\rho} (Ze^{\rho}) \ge 0$, using (d).
Integrating from $\rho$ to $\rho_{1}\left(\rho \in (0,\rho_{1}) \right)$,
shows that $Z(\rho)=0$ for all $\rho \in  (0,\rho_{1})$.
A similar argument using $\f{d}{d\rho} (Ze^{-\rho}) \le 0$
shows that $Z(\rho)=0$ for all $\rho \in  (\rho_{1},\infty)$.
Thus, if $Z(\rho_{1})=0$, then $Z \equiv 0$ on $I$. 
But this possibility is excluded by the non-degeneracy condition.
Therefore, $Z > 0$ on $(0,\infty)$. This shows that $Z$ does not
have isolated zeroes.

 $F$ is (strictly) monotonic increasing everywhere, since from (b)
 $\f{dF}{d\rho}=Z$ and $Z > 0$.  
If $F(\rho_{1}) \ge 0$ for some $\rho_{2}>\rho_{1}$, $F(\rho_{2})>0.$ 
So, for $\rho \in (\rho_{2},\infty)$ 
\[ \f{dA}{d\rho} = \f{F}{\rho} \ge \f{F(\rho_{2})}{\rho}.  \]
Integrating,
\[ A(\rho) \ge A(\rho_{2}) + F(\rho_{2})\ln \left(\f{\rho}{\rho_{2}}\right)  \]
and $A \rightarrow \infty$ as $\rho \rightarrow \infty$.

Also, from (a), $\f{d\chi}{d\rho} \geq A-1$, which implies, 
after integration, that $\chi \rightarrow \infty $ as $\rho \rightarrow \infty$.

Now on $(\rho_{2},\infty)$, $F$ is bounded below by $F(\rho_{2})$.
Let us assume that $F$ is bounded above, i.e. $0<F(\rho_{2})\leq F(\rho)\leq
\beta$, a constant.
Consider
\be
\f{d}{d \rho} \left(\f{Z(A-\cos\chi)}{A+1}\right) = \f{ZF(1+\cos\chi)}
{\rho(A+1)^{2}}.   \label{eq:deriv1}   
\ee
Now for $\rho_{2}$ large enough, $A>1$ on $(\rho_{2},\infty)$ and so
\[ \f{Z(A-\cos\chi)}{A+1} \;\;>\;\; c_{4} \;\;=\;\; \f{Z(\rho_{2}) 
\left( A(\rho_{2}) - \cos\chi(\rho_{2}) \right)} {A(\rho_{2})+1} > 0.  \]
Thus,
\be
Z(A-\cos\chi) > c_{4}(A+1).  \label{eq:zcos}
\ee
Now consider 
\[ V(\rho) = Z (A-\cos\chi) - \f{F^{2}}{2\rho} - \int^{\rho}_{\rho_{2}}
\f{F^{2}(\sigma)}{2(\sigma^{2}}\, d\sigma.  \] 
$\f{dV}{d\rho} = 0$, so $V(\rho)=V(\rho_{2})$, constant on $(\rho_{2},\infty)$.
But 
 \[V(\rho) \geq Z(A-\cos\chi)-\f{F(\rho_{2})^{2}}{2\rho}-\f{1}{2}F(\rho_{2})^{2}
\int^{\rho}_{\rho_{2}} \f{d\sigma}{(\sigma)^{2}},   \]
and so, using (\ref{eq:zcos}), since $A \rightarrow \infty$ as $\rho \rightarrow \infty$, then $V(\rho) \rightarrow \infty$ as $\rho \rightarrow \infty$.
Clearly $\rho_{2}$ can be chosen at a point where $V(\rho_{2})$ is finite.
We have a contradiction. $F$ cannot be bounded by a positive real number.
That is, $F \rightarrow \infty$ as $\rho \rightarrow \infty.$ 
   \item[(ii)] We assume there exists a constant $\eta$ s.t.
$F(\rho) < \eta < 0$ for all $\rho \in (0,\infty)$.
Then, since $\f{dA}{d\rho} = F/\rho$,
\[ A \leq A(\rho_{2}) + \eta \ln (\f{\rho}{\rho_{2}}),   \]
and so $A \rightarrow -\infty$ as $\rho \rightarrow \infty$.
Similarly $\chi \rightarrow -\infty$ as $\rho \rightarrow \infty$.

Now
\be
\f{d}{d \rho} \left(\f{Z(A-\cos\chi)}{A-1}\right) = \f{ZF\left(-1+\cos\chi)
\right)}{\rho (A-1)^{2}} \geq 0.  \label{eq:deriv2} 
\ee
Therefore
\[  \f{Z(A-\cos\chi)}{A-1} \geq c_{5} \;\;=\;\;\f{Z(\rho_{2})\left(A(\rho_{2})
-\cos\chi(\rho_{2})\right)}{A(\rho_{2})-1}  \geq 0,  \]
for $\rho_{2}$ large enough, since $A \rightarrow -\infty$ as $\rho 
\rightarrow \infty$. Therefore
\[ Z(A-\cos\chi) \;\;\leq\;\; c_{5}(A-1).  \]
Hence $Z(A-\cos\chi) \rightarrow -\infty$ as $\rho \rightarrow \infty$.

Now, using a similar argument to that in (i), using $V(\rho)$,
we have a contradiction. Hence, $F$ cannot be bound above by $\eta<0$.
But $\f{dF}{d\rho}=Z>0$,
thus $F \rightarrow 0$ as $\rho \rightarrow \infty$ (otherwise, we have
case (i)).
The proof of $Z \rightarrow 0$ as $\rho \rightarrow \infty$ is 
handled in a similar manner. The bounds on $A_{\infty}$ are also
easily established using arguments of this type. For example, assume
$A_{\infty} < -1$ and use (\ref{eq:deriv1}) and (\ref{eq:deriv2}).
\begin{flushright}  $\Box$  \end{flushright}
\end{enumerate}

\section{Numerical Solutions}
\subsection{Charged Wire with Two Dense Rings}

As stated in {\bf Lemma 2} $(ii)$, if $F<0$ on $I$, then $F \rightarrow 0$
as  $\rho \rightarrow\infty$.
To obtain a numerical solution of this type, we expand $\chi$ in $\f{1}{\rho}$
and solve for the coefficients of $\chi, A, F, Z$ as $\rho \rightarrow \infty$.
{\em There is only one such solution, the coefficients in the expansions
being uniquely determined}.
\bea
\chi  & = &\pi-3\f{1}{\rho}+\f{1}{6}\f{1}{\rho^{3}}-\f{9247}{67320}
                \f{1}{\rho^{5}} + \dots \nonumber  \\
A     & = & -1+\f{15}{2}\f{1}{\rho^{2}}-\f{35}{8}\f{1}{\rho^{4}}
          +\f{77425}{26928}\f{1}{\rho^{6}}+\dots \nonumber  \\
F     & = & -15\f{1}{\rho^{2}} + \f{35}{2}\f{1}{\rho^{4}}-\f{77425}{4488}
           \f{1}{\rho^{6}} + \dots \nonumber  \\
Z     & = & 30\f{1}{\rho^{3}}-70\f{1}{\rho^{5}}+\f{77425}{748}\f{1}{\rho^{7}} 
             + \dots \ .
\eea

Using these expressions, we evaluate our initial values (near infinity).
We used a multistep differential equation solver (from the 
NAG Fortran Library (Mark 16)\cite{for}), interfaced with {\bf MATLAB}
 (\cite{mat})).
The numerical solution is given in Figure~1.
\par
\begin{figure}
\par
\centerline{\hbox{
\psfig {figure=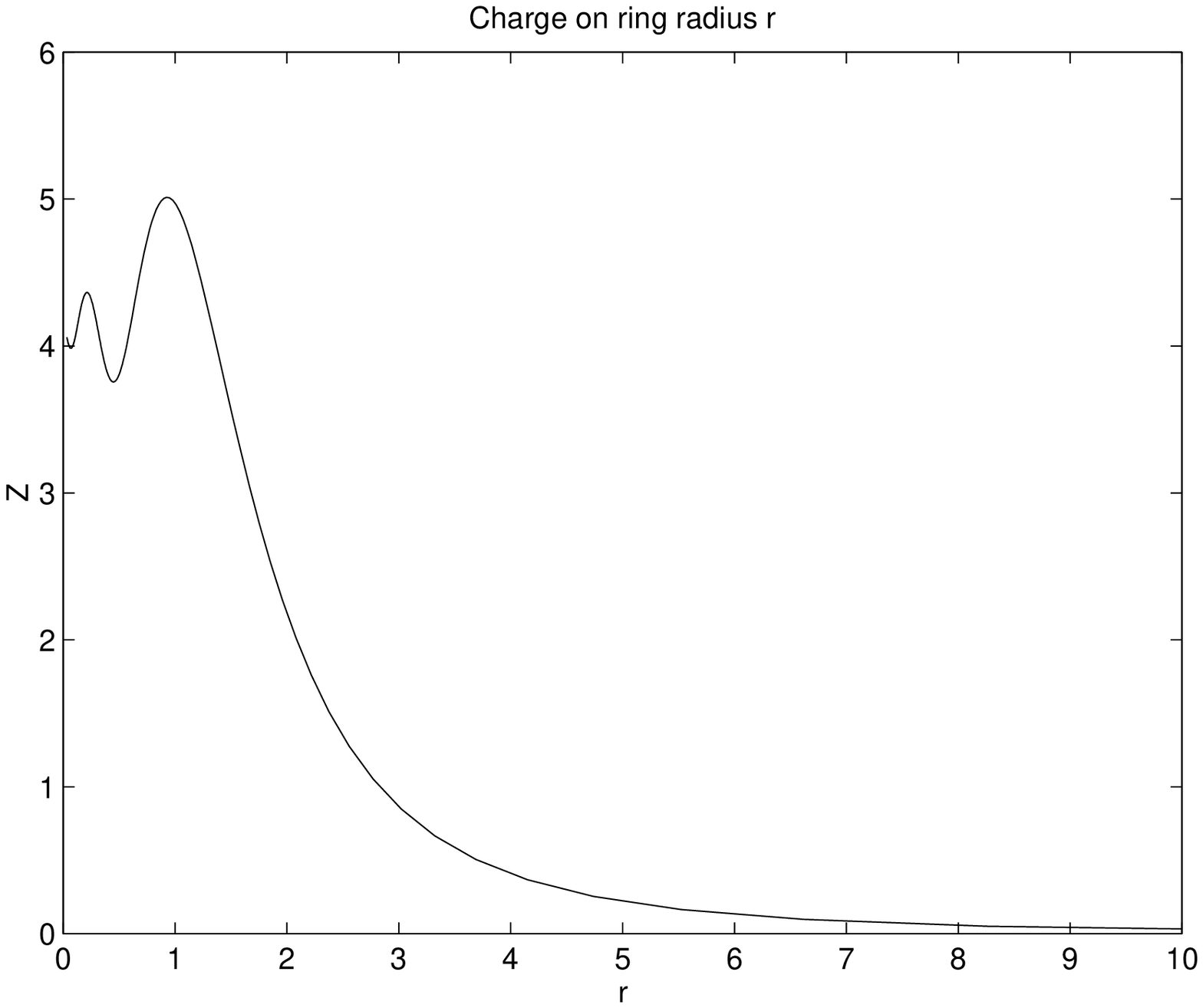,height=6cm}
\psfig {figure=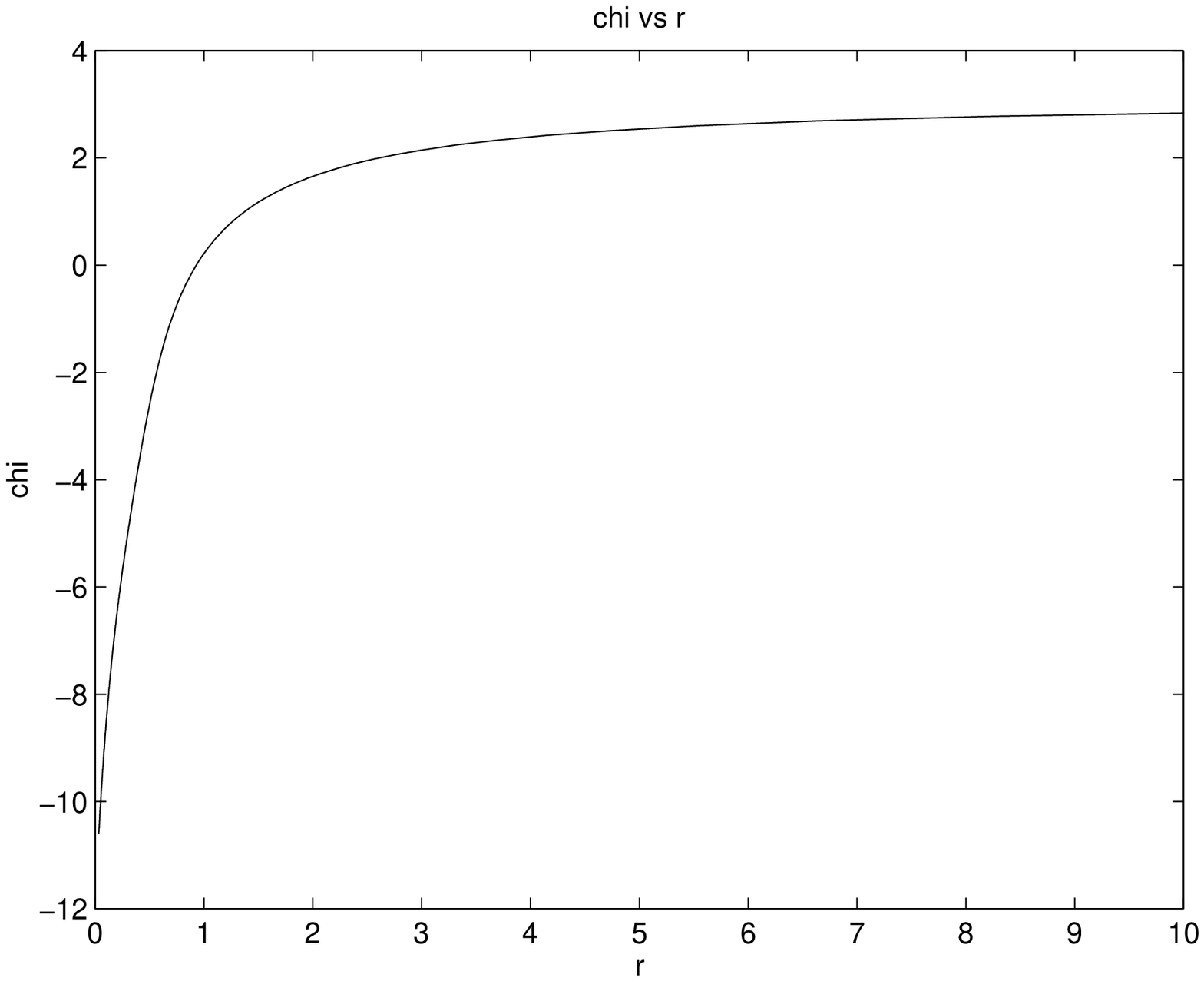,height=6cm}
}}
\par

\par
\centerline{\hbox{
\psfig {figure=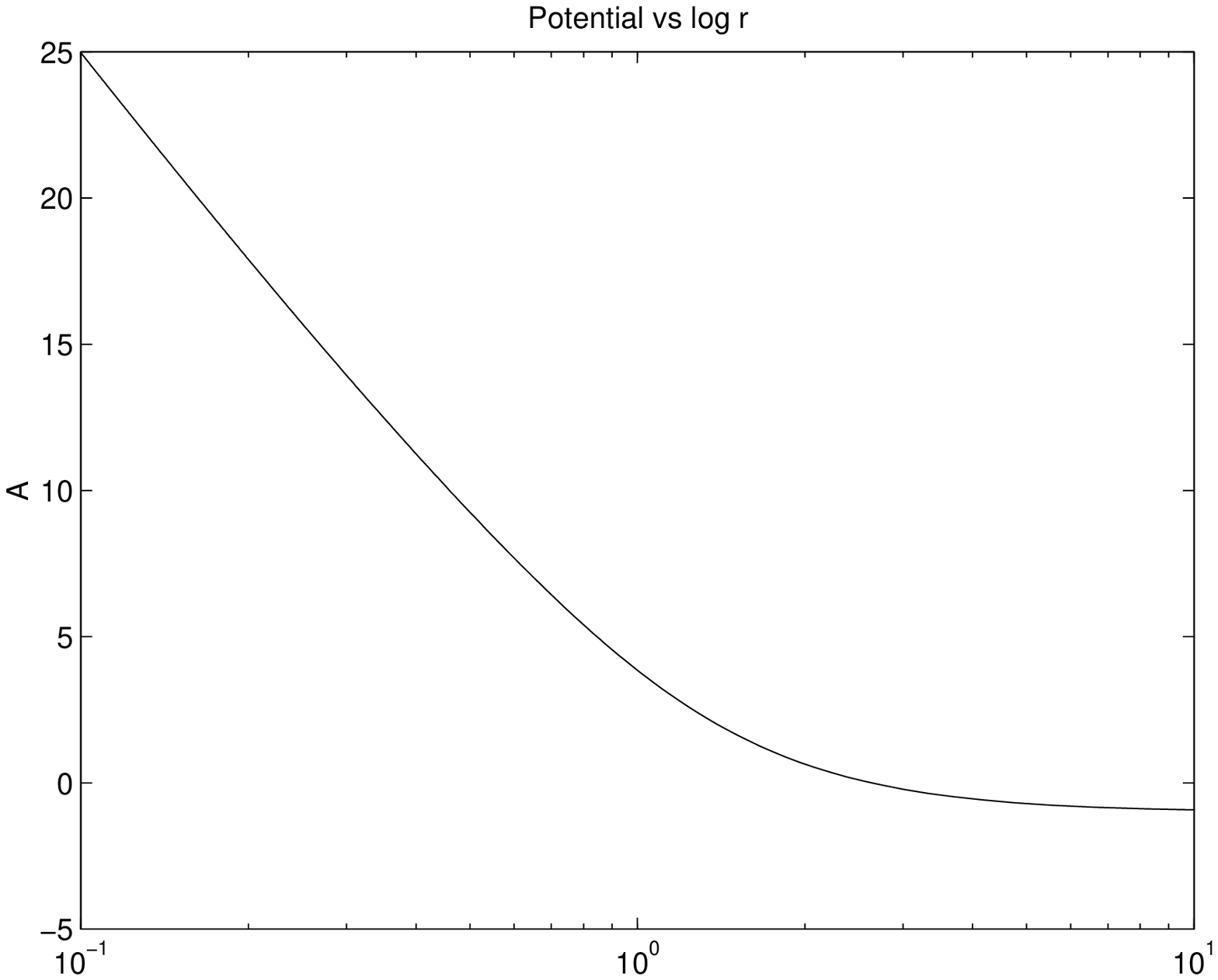,height=6cm}
\psfig {figure=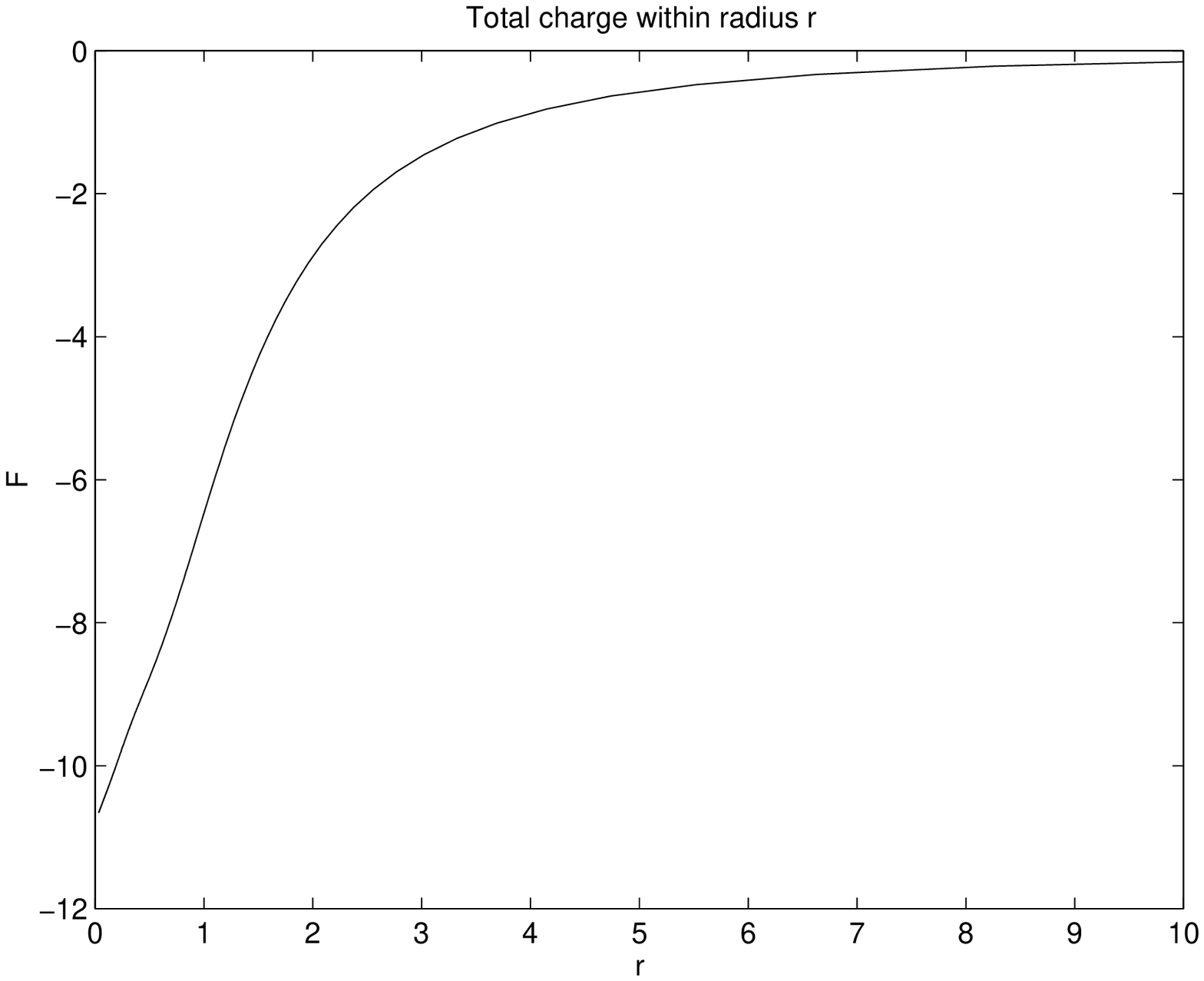,height=6cm}
}}
\caption{Numerical solutions to the static D-M equations, 
         with cylindrical symmetry}            

\par
\end{figure}

As stated in {\bf Lemma 1}, $\chi$ remains finite as $\rho \rightarrow 0$.
We can calculate the value of $\chi(0)$ numerically:
$\chi(0) = -12.17 $
noting that $\chi \rightarrow \pi$ as  $\rho \rightarrow \infty$.
From (d) we have
\[ \f{dZ}{d\rho} =0 \;\;\;\;\mbox{when}\;\;\;\;\chi=n \pi.  \]
From Figure~1 we see that $\chi$ is monotonic and bounded between $-4\pi$ and
 $\pi$.
$Z$ has four critical points for $\rho \in (0,\infty)$.
These occur successively as maxima and minima. $Z$
has two finite values of $\rho$, $\rho_{1}$ and $\rho_{2}$, say, for which
 the charge density is a local maximum.
See Figure~1. Numerically,
after converting to units, in which $\rho$ is measured in
$\lambda_{C} = \f{1}{m}$, the reduced Compton wavelength, 
$\rho_{1} \approx .12 \lambda_{C}$,
and $\rho_{2} \approx .46 \lambda_{C}$.
In the full three-dimensional picture, 
this corresponds to two densely charged rings around the $z$ axis.

Plotting $A$ against $\ln \rho$, in the inner region (inside the
dense rings) we obtain a linear plot, in keeping with {\bf Lemma 1}($ii$): 
\[ A \approx F(0)\ln \rho \;\;\;\;\mbox{for}\;\;\;\;\rho << \rho_{1}.     \]
This is just the standard cylindrically symmetric
solution to vacuum Maxwell equations -- the potential due to
a charged wire (along the $z$ axis). 
Note that, $F(0)= \f{146}{\lambda_{C}}$ corresponding
to a charge per Compton wavelength of $146e$.
Coupling the Maxwell and Dirac equations, then, has the effect of surrounding
the charged wire with two dense rings of charge.

\subsection{Comparison with the Linearized Theory}
In this section we consider, by way of comparison,
the solution to the Dirac equation
in an external field generated by a charged wire.
We now use the decoupled equations, i.e. the Dirac equation
with the vacuum Maxwell equation.

In the static cylindrically symmetric case, the solution to the
vacuum Maxwell equation is $A =c_{0} \ln \rho.$
Using the same notation as in the previous section, 
we choose $c_{0} = F(0) < 0$. 
We can now solve the Dirac equation, using this (external) potential.
Our equations are
\bea
(a) \;\;\;\;\;\;\f{d \chi}{d \rho} & = & -10.79 \ln \rho - \cos \chi 
  \nonumber  \\
(b) \;\;\;\;\;\;\f{d Z}{d \rho}   & = &  -Z \sin \chi.    \label{eq:2des} 
\eea
For comparison purposes, we solve these linearized equations with the
same initial conditions used for our earlier solution.
The numerical solution was obtained using the same methods
as in the previous section and are shown in Figure~2.
\par
\begin{figure}
\centerline{\hbox{
\psfig {figure=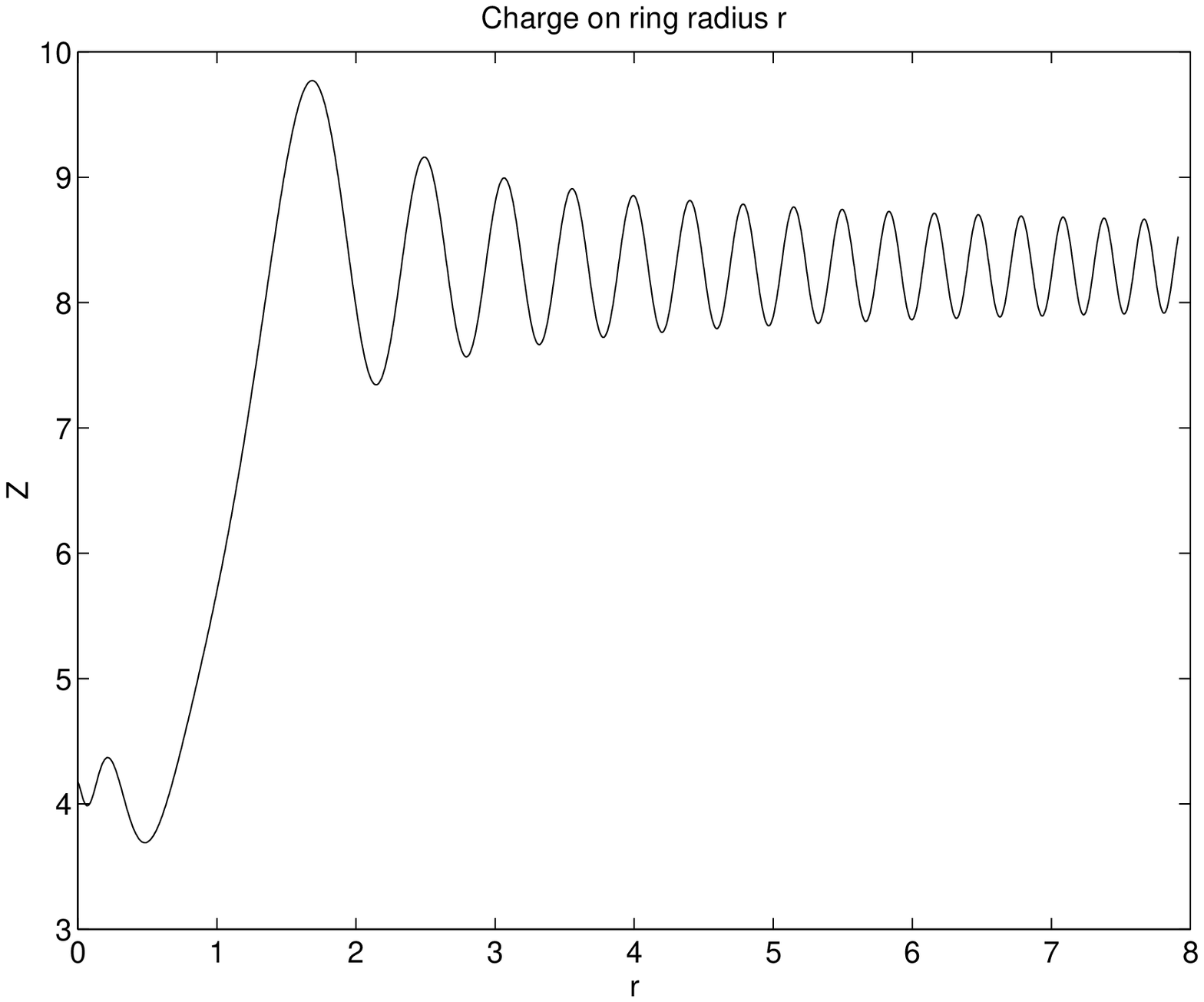,height=6cm}
\psfig {figure=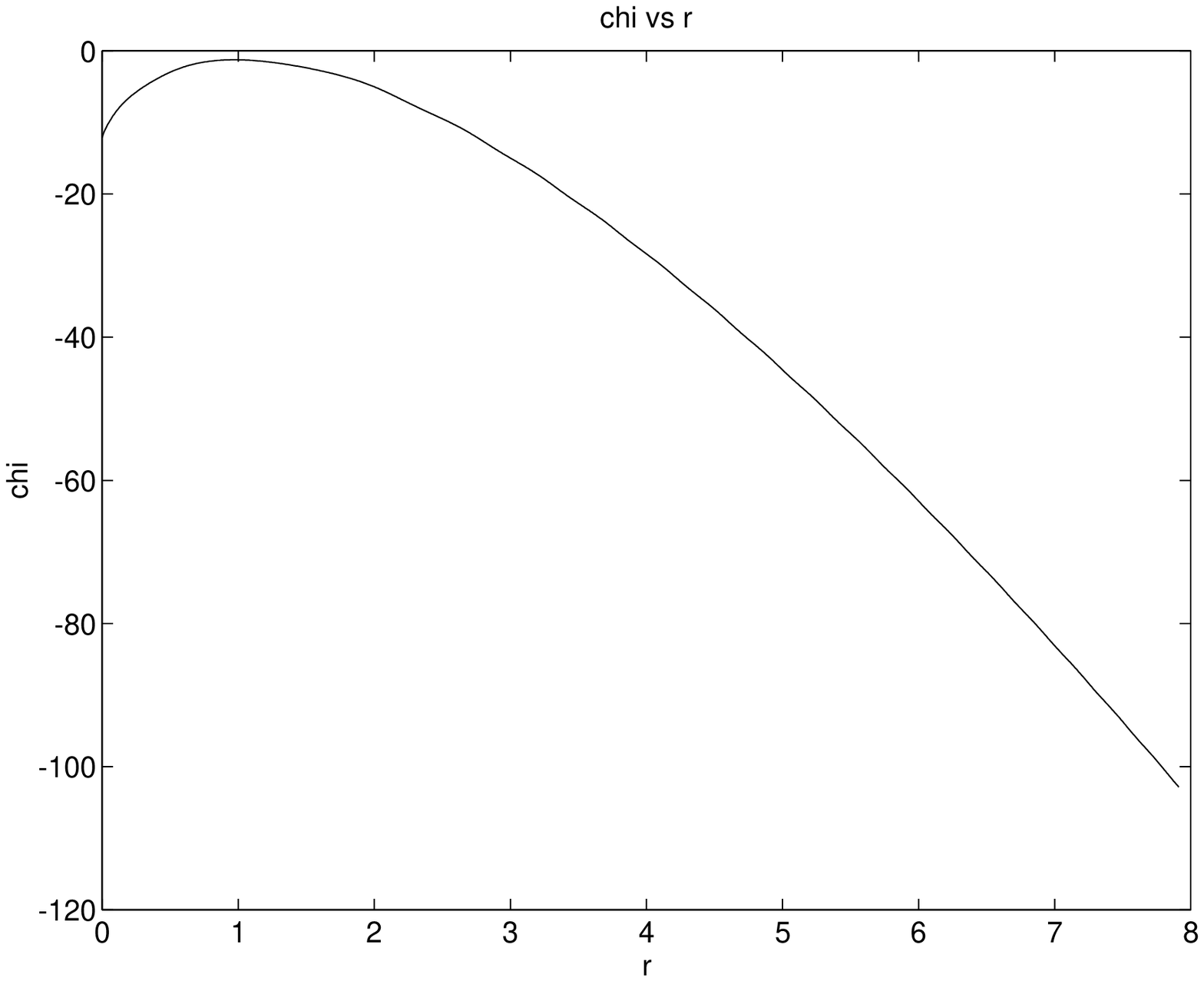,height=6cm}
}}
\caption{Numerical solution to the Dirac equation aound a charged
         wire, treated as an external field}
\end{figure}
\par

For small values of $\rho$ the current density, $Z$, behaves in
a similar manner (to the solution of the non-linearised D-M equations),
having the first peak at $\rho_{1} \approx .11\lambda_{C}$. 
As $\rho \rightarrow \infty$  the (logarithmic) potential  
is unbounded, in contrast to the nonlinearised case,
in which $-1 \le A_{\infty} \le 1$ (see {\bf Lemma 2}). 
It is clear from equation (\ref{eq:2des}$(a)$) that
$\chi \rightarrow -\infty$ as $\rho \rightarrow \infty$.
As such, (from \ref{eq:2des}$(b)$), there will be an infinite
number of oscillations in the charge density variable $Z$,
which, however, remains bounded.
We have lost the {\it localization} that was apparent
in the full D-M equations. 


\par
\par
\par
\par
\par
\begin{figure}
\par
\psfig {figure=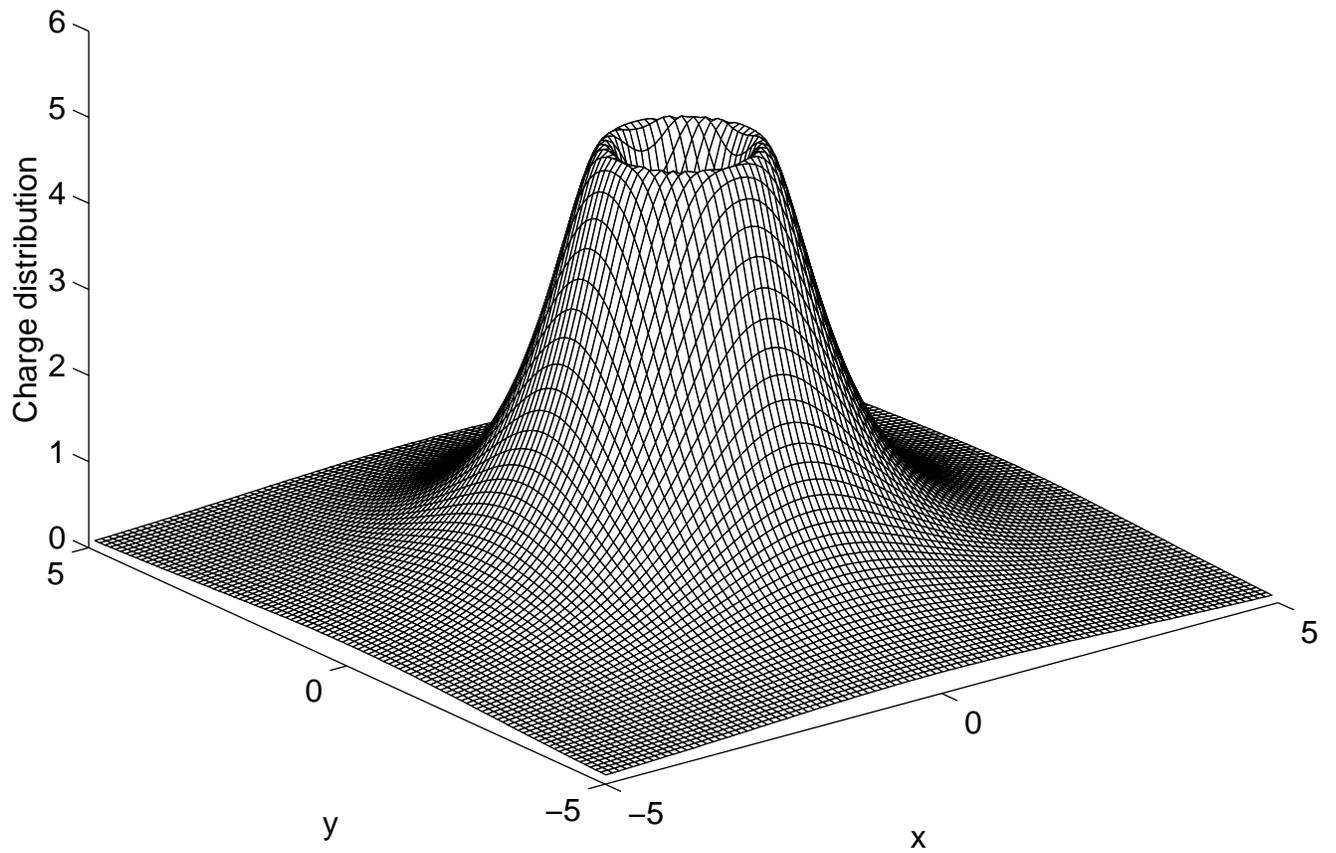}
\par
\caption{Charge distribution of the Dirac-Maxwell field with cylindrical
         symmetry evidencing a {\it highly localised} charge density
         around a central charged wire. Total charge (per unit length)
         is finite.}
\end{figure}
\par

\par
\begin{figure}
\par
\psfig {figure=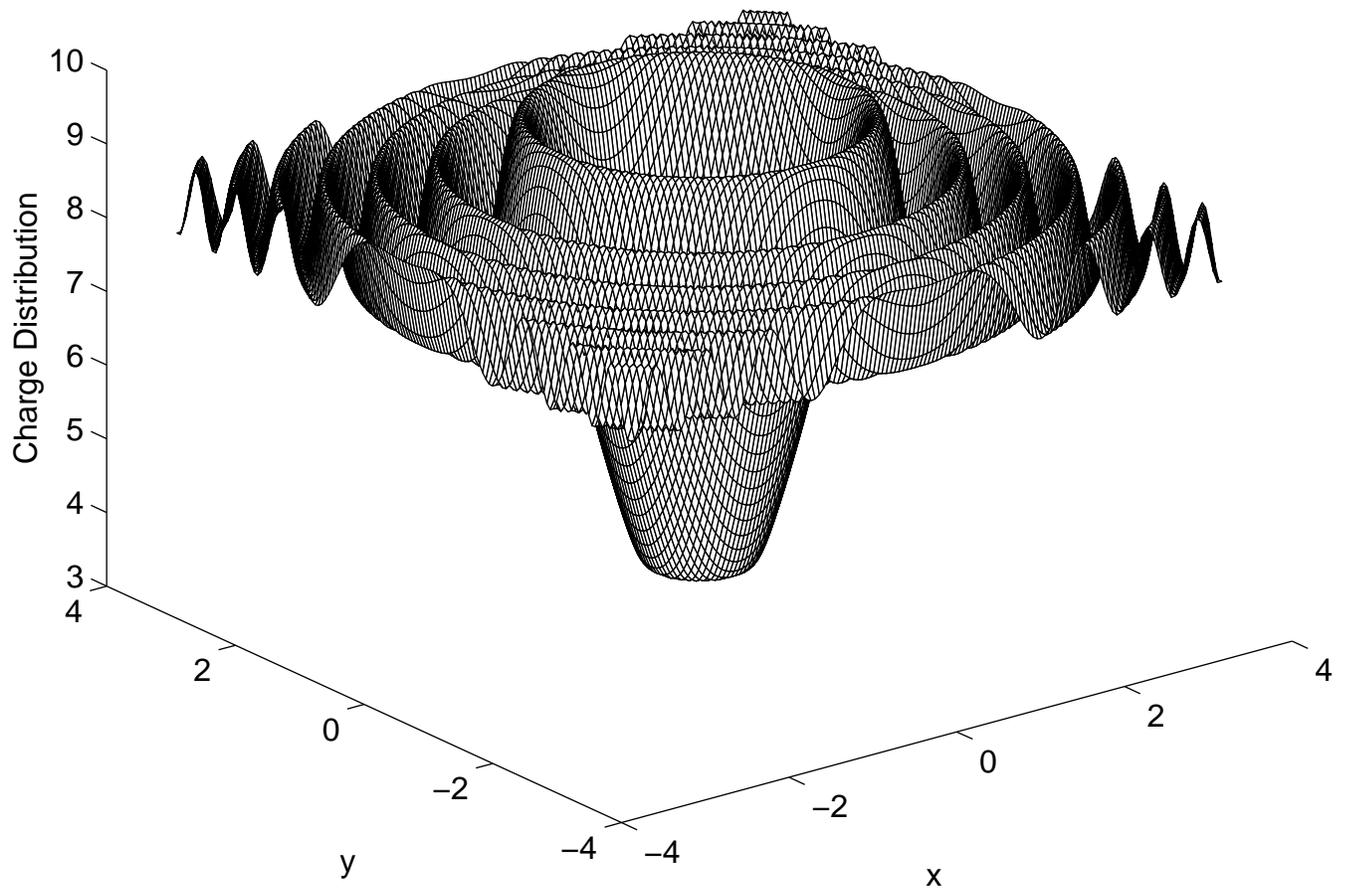}
\par
\caption{Charge distribution of the Dirac equation surrounding
a charged wire, treated as an external field. No localisation
of the charge distribution is apparent. The total charge (per unit
length) is unbounded.  }
\end{figure}
\par

\subsection{Unbounded Solutions}
 
As stated in {\bf Lemma 1}$(ii)$, if $F(0)\neq 0$ then $A$ is unbounded
as $\rho \rightarrow 0$.
We can also look for solutions where $F(0)=0$.
{\bf Lemma 2} then tells us that $\chi, A, F \rightarrow \infty$
as $\rho \rightarrow \infty$.
Since $F(\rho)$ is the charge within a radius $\rho$,
these solutions are of less interest, as the total charge,
$F(\infty)$ is  necessarily unbounded.

To find a numerical solution of this type, we expand $\chi$ into a Taylor
Series in $\rho$ and solve for the coefficients.

As an example of this type of solution we set $A(0)=1$
and $\chi_{0}=0$. The first few terms of these and the remaining
variables are given below.
\bea
\chi  & = & \f{1}{100}\rho+\f{1}{100}\rho^{2}+\f{1}{60000}\rho^{3}
  +\f{1}{45000}\rho^{4} + \dots \nonumber\\
A     & = & \f{101}{100}+ \f{1}{50}\rho-\f{1}{90000}\rho^{3}-\f{1}{240000}
        \rho^{4} + \dots \nonumber  \\
F     & = & \f{1}{50}\rho-\f{1}{30000}\rho^{3}-\f{1}{60000}\rho^{4} + \dots
       \nonumber  \\
Z     & = & \f{1}{50}-\f{1}{10000}\rho^{2}-\f{1}{15000}\rho^{3} 
             +\f{67}{4000000000}\rho^{4} + \dots \ .
\eea

In Figure~5 we show the solution which has these initial values.
\par
\begin{figure}
\centerline{\hbox{
\psfig {figure=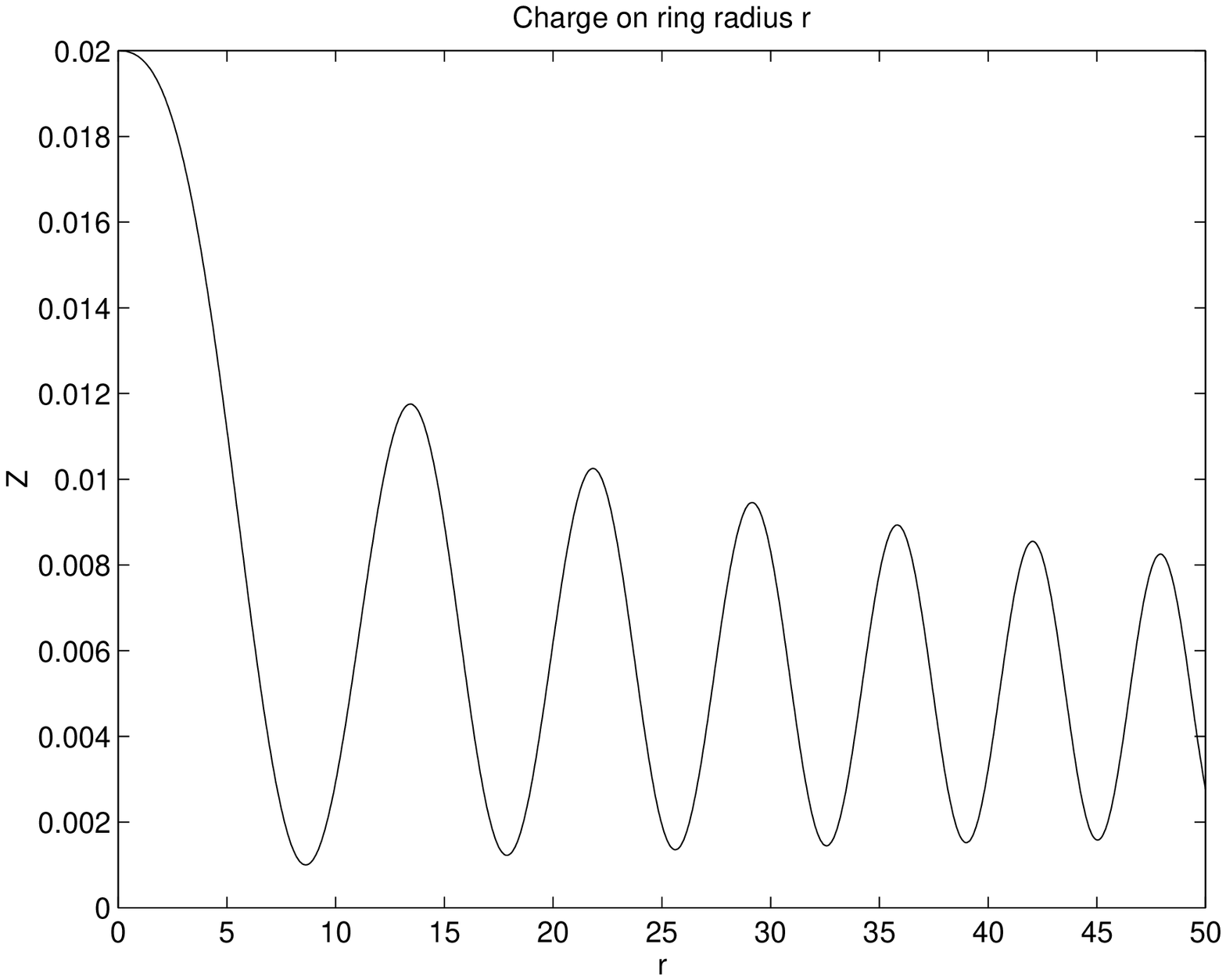,height=6cm}
\psfig {figure=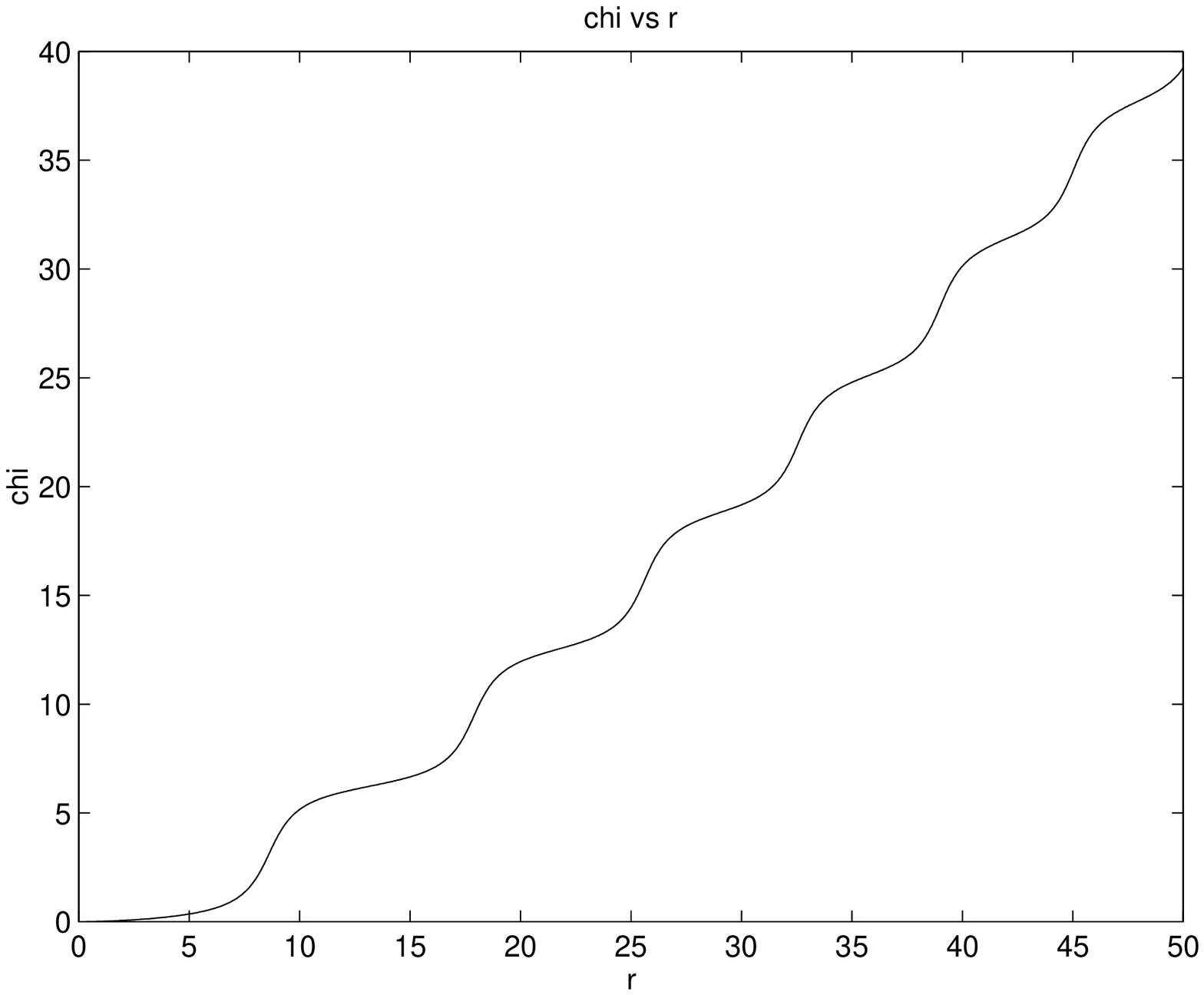,height=6cm}
}}
\par
\par \centerline{\hbox{
\psfig {figure=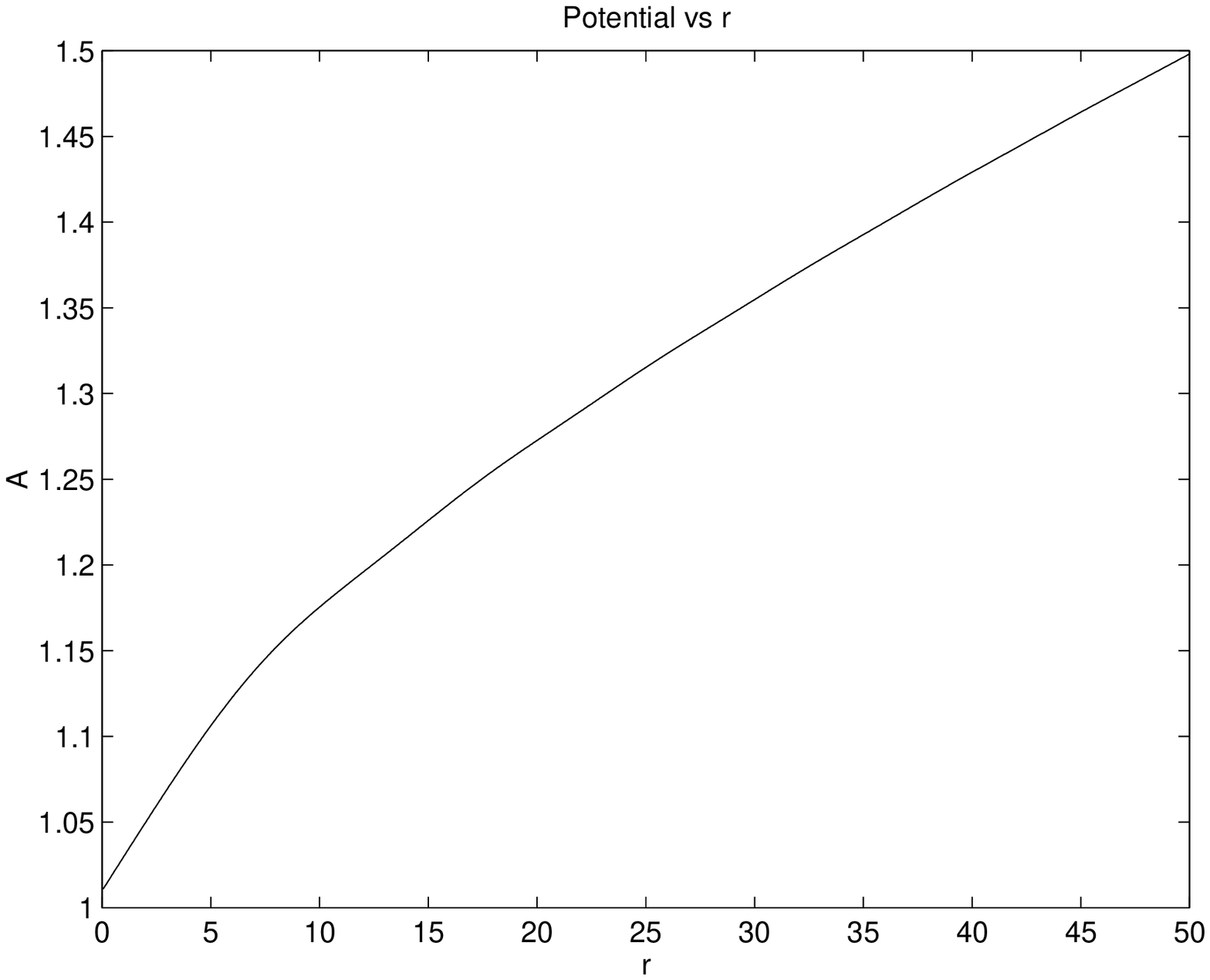,height=6cm}
\psfig {figure=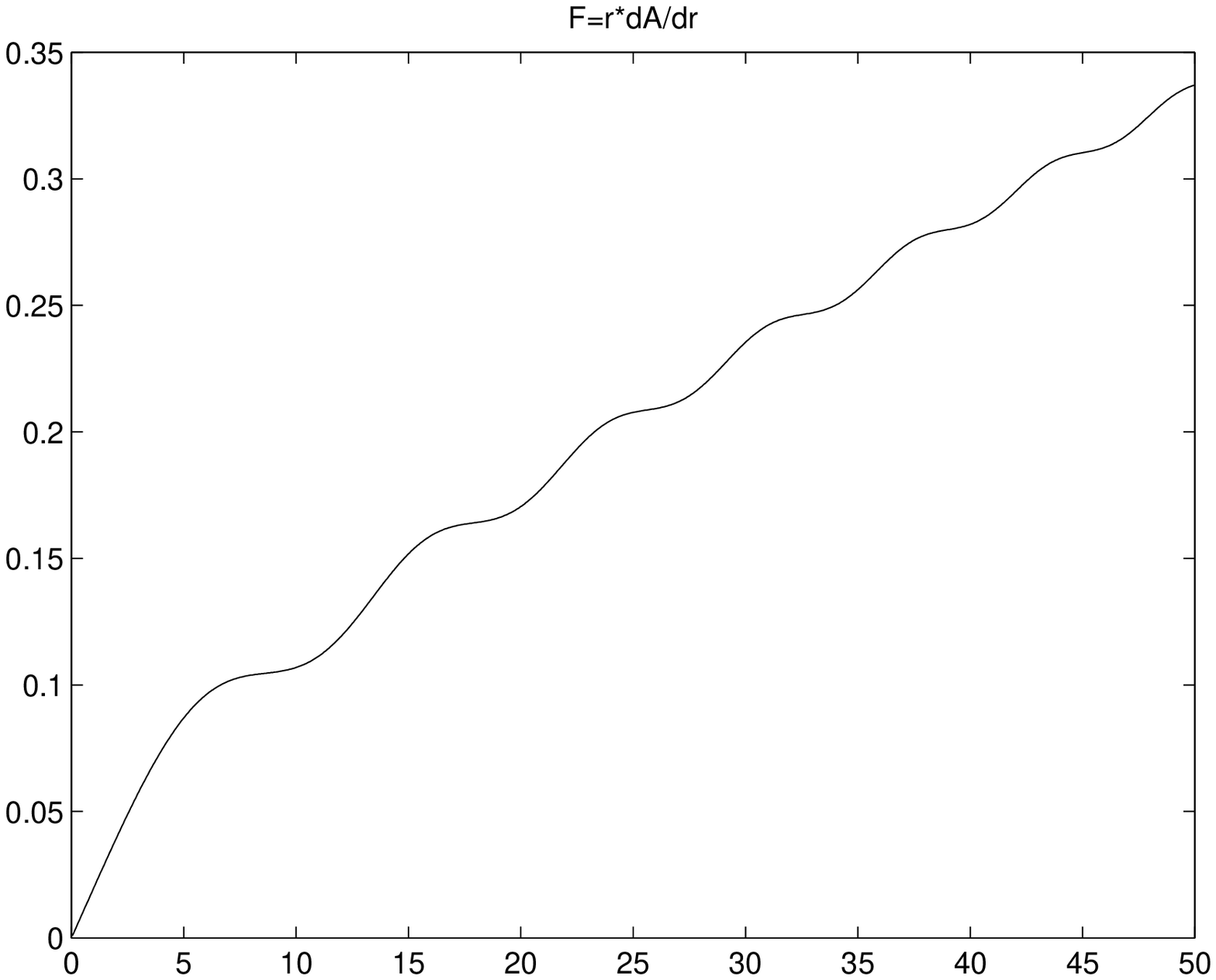,height=6cm}
}}
\caption{Numerical solutions to the D-M equations, with $F(0)=0$.}
\end{figure}
\par
\section{Conclusions}
An examination of the behaviour of the solutions to the static
Dirac-Maxwell Equations with cylindrical symmetry shows that, in
one class of solution, we have a highly localised charge density
around a central charged ``wire''. The total charge (per unit
length of the ``wire'') is finite. 

When the equations are decoupled (by ignoring the effect of
the ``self-field'' upon the Maxwell field) {\em all localisation
is lost}. The total charge (per unit length) of the Dirac field is,
 in this case, unbounded.

Although the cylindrical case is of limited interest physically,
the resulting o.d.e.s are amenable to a detailed descriptive
analysis which corrorborates the numerical results.

The same localisation due to the inclusion of a nonlinear
coupling, will be more interesting in the axially
symmetric case, an analysis of which will be in a forthcoming publication.
\section{Acknowledgements}
The first named author was partially supported
by an Australian Postgraduate Award.

\end{document}